\documentclass[11pt,twoside]{article}

\usepackage{asp2006_ecl}

\begin{document}
\title{Panel discussion I: Star formation in galaxies: how do we continue?}
\author{Johan H. Knapen (Reporter)}
\affil{Instituto de Astrof\'\i sica de Canarias, E-38200 La Laguna, Spain}

\begin{abstract} %%% Abstract to run on from here.

This is the written account of the first of two panel discussions, on
{\it Star formation in galaxies: how do we continue?} The chair of the
panel was Phil James, and panel members were John Beckman, Torsten
B\"oker, Daniela Calzetti, Angeles D\'\i az, and Rob Kennicutt. The
panel and audience discussed the four following questions: 1) What are
the most critically needed techniques to give accurate measurements of
total rates and efficiencies of star formation?  2) Do we understand
the form of the initial mass function and its variation as a function
of redshift and environment?  3) Are there multiple modes of star
formation in galaxies (bulge vs. disk, burst vs. continuous) or does
the Schmidt law explain everything?  4) How do we bring together our
understanding of star formation in our Galaxy and in external systems?

\end{abstract}

%%% MAIN BODY OF TEXT GOES HERE. CONSULT "INSTRUCTIONS FOR AUTHORS USING
%%% LATEX2E MARKUP", SECTIONS 2.3-2.6 FOR HELP WITH EQUATIONS, FIGURES,
%%% AND TABLES.

\medskip

{\sc James:} (Introduced the panel members, then introduced the four
questions, and asked John Beckman to give his opinions on the first:
What are the most critically needed techniques to give accurate
measurements of total rates and efficiencies of star formation (SF)? )

\medskip

{\sc Beckman:} Rob (Kennicutt) and Daniela (Calzetti) will have more
to say on this topic but I will make some remarks to start the
discussion. One question is on missing photons: which fraction of the
SF rate (SFR) might be missed due to escaping photons when we observe
a whole galaxy (assuming that the initial mass function [IMF] issue is
covered). We know that photons can easily leak out of H{\sc ii}
regions, but most will get soaked up somewhere in the galaxy so the
global SFR should take them into account anyway. But there is a
caveat. In grand design galaxies in particular, a large fraction, up
to 50\%, of the emission may be diffuse. This should not be missed but
the diffuse emission can easily fall below the detection
threshold. Locally this is most probably not a problem, but it might
well be more significant at higher redshifts.

\medskip

{\sc D\'\i az:} I would like to stress the importance of high accuracy
in measurements. A second point is that we then need similar accuracy
in the tools we subsequently use for analysis. We need to know and
quantify the influence of, e.g., leaking photons, but also second
order effects which we must take into account as the observations
become better. We also need to analyze the scatter which may well
carry physical information.

\medskip

{\sc Kennicutt:} To me, the key is the issue of efficiencies. Thanks
to the multi-wavelength revolution in observations over the last few
years we now have a lot of information on many different issues. But I
worry about our knowledge of the interstellar medium, about molecular
masses involved, and about cloud cores. New facilities like CARMA and
ALMA are coming, but we need experts to say where improvement is most
needed.

\medskip

{\sc B\"oker:} A main outstanding question is that of recent versus
current SF. Even H$\alpha$ gives information on the ``past'' SF, and
to observe the current SF we need sub-millimeter observations. This is
not yet being incorporated into the discussion on SFRs.

\medskip

{\sc Calzetti:} Among the advances that new millimeter telescopes and
instruments (ALMA, LMT [Large Millimeter Telescope], CARMA) will bring
is the possibility of detecting CO emission in the interarm regions of
galaxies, another under-explored region of the parameter space of
galactic environments.

\medskip

{\sc James:} Thank you panel members. I now invite comments from the
audience.

\medskip

{\sc Krumholz:} I would like to comment on questions (1) and (4), and
stress that we need to worry most about, and clarify, the efficiency
of SF. In a Galactic context, the term efficiency is used to describe
the fraction of mass which is transformed into stars over the lifetime
of a cloud. But in an extragalactic context, efficiency is used to
denote the SFR over gas mass, which is an instantaneous measure and
not one over a lifetime. My comment, then, is that we should rather
{\it not} use the term efficiency, or define carefully whether we are
talking about instantaneous efficiency or over a lifetime.

\medskip

{\sc Hensler:} I agree with Mark (Krumholz) but let me add that a
galaxy is not a closed box, and so in addition we need to consider
how, e.g., infall influences the efficiency.

\medskip

{\sc Sheth:} The new CARMA telescope is now working, and early work on
M51 shows that it works very well. But rather than concentrating on
HCN, we should use 1\,mm continuum which is where the improvements in
CARMA are the most significant compared to OVRO/BIMA and other
observatories.  The 1\,mm continuum would trace the dense dusty
regions in galaxies.

\medskip

{\sc Evans:} A worry is whether even with ALMA we will be able to sort
out observational issues to obtain molecular gas masses to a precision
better than a factor of three. In particular the calibration of CO
measurements is difficult and we need simulations of molecular clouds
in galaxies to understand what the observations mean and what their
real uncertainties are.

\medskip

{\sc Churchwell:} IR observations are sensitive to the most massive
young stellar objects (YSOs), whereas X-ray observations are sensitive
to magnetic activity in lower mass YSOs, so together one has a chance
to determine the IMF for the whole stellar mass range within a given
cluster. It's the lower mass range that has historically been the most
difficult to establish and since most of the mass is in lower mass
stars, this is the most important part of the IMF that needs to be
determined with good precision. My student is trying to use the
GLIMPSE survey to determine the current global SF rate in the Galaxy,
but to do this he has to determine the IMF, especially for lower mass
YSOs, for various environments in the disk of our Galaxy. Our hope is
that X-ray data from {\it Chandra} and other X-ray satellites will
provide the tool to do this.

\medskip

{\sc Lehnert:} I would like to address the differences that we heard
on Monday between the results on SF modeling, from Mark (Krumholz)
and Ian (Bonnell). We need to know how SF happens in different
environments within and outside the Galaxy, and the key point is how
do we approach this.

\medskip

{\sc Krumholz:} The difference is mainly due to a limitation in the
CPU power we can use. Ian (Bonnell) can model larger volumes because
he doesn't include the effects of radiation.

\medskip

{\sc Lehnert:} ...but are you sure you know the physics?

\medskip

{\sc Krumholz:} The limitation is in the code and in the CPU, not in
the physics.

\medskip

{\sc Kroupa:} We've heard earlier in the week (in the talk by Alves)
that the stellar IMF shape is already found in the mass distribution
of the cloud cores. But theoretical work seems to be wrong, for
instance I quote the failure of Klessen, Spaans \& Jappsen (2007)
to reproduce the mass function of the Arches cluster. This is very
worrying indeed, and we need more work on the theoretical side.

\medskip

{\sc Bonnell:} The Klessen et al. result of a top-heavy IMF depends
primarily on their input assumptions of the dust physics and
background radiation field. I would not misconstrue their result as
implying we do not understand the dynamics of the star formation
process, but rather that there remain many unknowns that are still
input into the numerical simulations.

\medskip

{\sc Alves:} I propose the use of a combination of the extinction
technique, which is easy and works well in local clouds, with the
study of clouds at larger distances as described by Churchwell on
Monday. The prospects of this combination are good, but it won't be
easy.

\medskip

{\sc Evans:} I totally agree. The cores we see in local clouds go up
to $2-3\,M_\odot$ in mass and have nothing to do with massive SF. Up
to now we cannot resolve the much larger cores, and even with ALMA
this will be very tough.

\medskip

{\sc James:} I propose that we move on to point (3): Are there
multiple modes of SF in galaxies (bulge vs. disk, burst
vs. continuous) or does the Schmidt law explain everything? Let's hear
from our panel members.

\medskip

{\sc Beckman:} Before going to point (3) I would like to add something
to the discussion on the previous point, namely that the
correspondence between the IMFs as derived from cloud cores and stars
is almost too good to be true - and let's hope it is indeed true. This
makes the theorists' life much easier, all they will have to do is
scale...  I think the measured turnover in the stellar IMF to low
masses, plus the turnover in the mass function of the cloud cores in
J\~{o}ao's (Alves) work is clearly one of the key novelties so far in
this meeting.

\medskip

{\sc James:} On to point (3) then - we should discuss whether the
Schmidt-Kennicutt law is really the last word, or whether it may in
some sense be holding back new developments in this area, and whether
we should thus look to move on to a more physically grounded approach
(while still citing Rob (Kennicutt) appropriately)?

\medskip

{\sc Beckman:} Rob?

\medskip

{\sc Kennicutt:} I'm thinking on my feet here... In yesterday's talks
we heard about SF in nuclei, we know now that this occurs over scales
of hundreds of parsec, and is involved in the formation of bulges. We
have heard of central SFRs of up to 1000\,$M_\odot$\,yr$^{-1}$. It is
remarkable that one law can cover all this range of SF, from the very
low-mass regime to the extremes just mentioned. I wonder whether we
have any evidence that there is more than one mode, or more than one
phenomenon. Does the single Schmidt law reflect a single universal
mode of SF, or could there be two or more modes that happen to follow
the same dependence of SFR and gas surface densities?  I guess the
answer in this respect is very much like we have discussed already
with regards to the IMF: we don't see any evidence for a change, but
we should be astounded if indeed there weren't any.

\medskip

{\sc Beckman:} Suggestions on whether there is evidence from abundance
ratios for more than one mode of SF come and go -- does anyone know
what the current state of affairs is?

\medskip

{\sc B\"oker:} The SF mode that produced bulges in the early Universe
must be different from the currently dominant one: a stellar mass in
the order $10^9-10^{10}\,M_\odot$ was formed in a very short time.

\medskip

{\sc Kennicutt:} Or for example how did a globular cluster like M13 or
Omega Cen form?  Have we seen anything at this conference that is
relevant to SF in that context?

\medskip

{\sc D\'\i az:} Considering clusters of 1000\,$M_\odot$ (not
1000\,$M_\odot$\,yr$^{-1}$!), do we have there a violent mode of SF,
such as the one that Rosa (Gonz\'alez-Delgado) talked about? It cannot
simply be adding up lots of little SF events, this doesn't seem to
work so we must have another mode of SF.

\medskip

{\sc Calzetti:} I would like to present a concrete example on how
observations can sometime be misleading in the interpretation of the
modes of SF.  Back in the mid-90's, analysis of UV images of nearby
starburst galaxies led to the discovery that only a small fraction
(about 20\% or less) of the UV light was associated with clearly
identifiable stellar clusters; most of the UV light ($\sim$80\%) was
associated with a diffuse stellar component.  During that period, the
hypothesis was advanced that SF had two ``flavors'' (modes): a cluster
mode and a ``diffuse'' mode. Subsequent studies, over the course of
the following 10 years, have shown that the diffuse UV component is
likely the ``product'' of the dissolution of the stellar clusters over
timescales of about 10--20\,Myr; once the clusters have evaporated,
the UV light will appear to be originating from a diffuse stellar
population. Hence, there appears to be one single mode of SF in
starbursts, the ``cluster'' mode.

\medskip

{\sc James:} Let me restate the question posed by John (Beckman): is
there any evidence from stellar abundance ratios for the existence of
different modes of SF?

\medskip

{\sc Silk:} In this respect, we remain biased by the local
Universe. Even if we can cope with objects like M13, we continue to
have big problems at high redshifts.

\medskip

{\sc Bastian:} Looking at smaller scales, we know from a decade of
star cluster research with the {\it HST} that there are {\it not}
different modes.  The relations between the SFR and the number of
clusters present, and similarly the relation between the SFR and the
mass of the most massive cluster in the galaxy argue strongly that
``starbursts'' and quiescent galaxies lie on the same continuum.  In
addition, work done by myself and also by Bruce Elmegreen has shown
that star clusters themselves are simply part of a continous
distribution of SF in galaxies, from sub-parsec to kpc scales.

\medskip

{\sc B\"oker:} This is a valid point, but it refers to the very local
Universe, and to the current SF. The past may have been very
different.

\medskip

{\sc Nesvadba:} We looked at the SF in detail in one galaxy at $z\sim
2.6$, as close as we can at present at high redshift (Nesvadba et
al. 2007). We found a SFR of $\sim500\,M_{\odot}$\,yr$^{-1}$, and no
evidence for an AGN contribution (which includes the X-ray). The SF
efficiency we found is very similar to that found at low redshifts,
which implies that the SF properties are very similar. The same is in
fact true for the SF-related feedback. There are obviously limitations
working at these redshifts, but empirically, we have not found
alarmingly large discrepancies.

\medskip

{\sc Krumholz:} A comment on the differences in the IMF. If the
expected number of low-mass stars were present in the ring around the
Galactic center (given the observed number of B-stars) we should have
seen X-ray emission from them, but haven't.

\medskip

{\sc Hammer:} We don't even know if there is a Schmidt law at
$z\sim0.6$. Indeed the gas distribution has a big effect on this law,
and we know that there is much more gas at a redshift of unity. How is
this gas distributed? I am not sure that question (3) is relevant.

\medskip

{\sc Capuzzo-Dolcetta:} May I stress that even after the talk by
Kennicutt it is not clear what the Schmidt law actually means
physically. As it stands, it is a phenomenological law.

\medskip

{\sc Edmunds:} I wonder why we must go on about variations in the
IMF.  The  point  is that it is fair enough that the IMF
changes in strange places like the Galactic center or the metal-poor
early Universe. But we don't have to change the overall IMF -- this is
great science! 

\medskip

{\sc Lehnert:} The Schmidt law has been alluded to as an empirical
one, but the question is what keeps the SF in-efficient.  It may be
related to the effect of shear (cf. Mark Krumholz) due to differential
rotation in the large scale velocity fields of galaxies, which will be
much lower in the central parts with solid body rotation. So, what
will control SF in these regions? And why do spirals look young, but
disks look old?

\medskip

{\sc Balcells:} Let me put on a theorists' hat. I would then want the
SF to depend on dynamical parameters, such as the velocity
dispersion. Gas clouds should collapse in different ways, and I would
like to have a parameter that prescribes when the gas can become
molecular. For instance, in ellipticals the gas is too hot so it won't
form stars. Similarly there is gas at the centers of spirals but it
cannot collapse. All this should really go into the Schmidt law, but
in that case will it still work?

\medskip

{\sc Kennicutt:} There are examples of that, e.g., the Silk law (which
relates the SFR density to the ratio of gas density and dynamical
time, and thus identifies the number of ``compression events'', or the
number of times the gas goes through the spiral arms). We have tested
this, but there is no clear signature.  For instance, in disks the SF
declines exponentially with radius, but the velocity dispersion
profile is flat with radius. So yes we do need to move to a physical
basis of the Schmidt law, but Occam's razor tells us that density
alone does the trick.

\medskip

{\sc Sarzi:} About star formation in early-type galaxies, it is
interesting to notice that the objects that for sure experienced
recent SF all have gas emission characterized by very small velocity
dispersion, independent of the apparent ionization of the gas, as
traced by the [O{\sc iii}]/H$\beta$ line ratio.

\medskip

{\sc Krumholz:} In response to Marc Balcells: there is no lack of
theoretical ideas, quite to the contrary, there are perhaps more
theories than theorists. There are papers by, e.g., Klessen, Silk, and
myself on how the Schmidt law may be related to the underlying
physics.

\medskip

{\sc Silk:} I agree that the Schmidt law is a very good fit and is a
great result, but the more interesting physics is in the scatter.

\medskip

{\sc D\'\i az:} Indeed, we should perhaps concentrate on those
objects that do not fit the Schmidt law and then try to find the
physical drivers that made them deviate.

\medskip

{\sc Calzetti:} We studied M83 in detail to test this possibility.  If
the interesting environments to study are those where SF deviates from
the Schmidt-Kennicutt law, then the outer regions of galaxies need to
be excluded from those studies. The combination of recent data in
H{\sc i} (from the VLA, data obtained by Fabian Walter), mid-infrared
(from {\it Spitzer}), and UV (from {\it GALEX}) of the outer regions
of M83 suggests that the UV emission in those extreme environments is
associated with SF that is compatible with the Schmidt-Kennicutt law.
Those regions are located at a distance from the center of about four
times the H$\alpha$ cut-off radius of M83.

\medskip

{\sc Beckman:} Partially in answer to Mike Edmunds' point: the
interest lies in trying to identify the objects that do not fit the
relation, and then to find the physics drivers for them.

\medskip

{\sc James:} I wonder if there are any further comments, because we
need to finish in five minutes. Would anyone from our panel perhaps
wish to comment on the fourth question: How do we bring together our
understanding of SF in our Galaxy and in external systems?

\medskip

{\sc Calzetti:} When I formulated question number four last night,
this is what I had in mind.  We have seen during the course of this
conference that there is a direct (linear) correlation between SFR and
dense cores. I believe this implies that we are moving the problem
from having to explain physically the Schmidt-Kennicutt law in its
original formulation, to having to explain the correlation between the
density of dense cores and the total gas density.

\medskip

{\sc Kennicutt:} If I refer back to the introduction of Neal's (Evans)
talk, I think we need to come together in our studies of the small and
the large scales. Perhaps the key to tying the Schmidt law to the
underlying physics lies in detailed studies of SF at small scales. So
Neal, let me put you on the spot: what should we observe?

\medskip

{\sc Evans:} We need to speak, first of all, a common language, use a
common terminology, especially on issues of efficiency and
timescales. There is a difference between whether, and how fast, you
make stars, and on large scales SF tends to be slow. Remember that the
actual process of SF occurs in dense cores, which are two steps
removed from the scales where the Kennicutt-Schmidt relations
operate. First, you need to understand what controls the formation of
molecular clouds, which is probably best studied with very high
resolution and sensitivity (ALMA and CARMA) in nearby galaxies. Then
you need to understand what controls the fraction of a molecular cloud
that turns into dense cores (a few percent in local molecular clouds,
but maybe very high in extreme starbursts). This is probably best
studied in our own Galaxy, but it requires a careful combination of
observations with models.  Those two approaches together may bridge
the gap.

\medskip

{\sc Sheth:} May I take the opportunity to advertise a conference
organized by the Spitzer team in Pasadena this autumn, where we plan
to bring together these two communities, and where we hope to start
answering these questions.

\medskip

{\sc James:} Our time has run out, so may I ask John Beckman to make a
closing comment?

\medskip

{\sc Beckman:} It is great to see more people getting together to
discuss these themes, the more people together, the more we think we
know. There are many new observational results, for instance
multi-wavelength determinations of SFRs using {\it GALEX}, {\it
Spitzer} and other facilities, and also on gas and its detailed
structure. So the observations are getting better. There still seem to
be too many theories so what we must perhaps think of are
observational results that cannot be reproduced by any theories.

\medskip

{\sc James:} With those words we finish this panel discussion. Let me
thank our panel members, and the members of the audience.

\end{document}